%% file: main.tex
\documentclass{article} 
\usepackage{iclr2026_conference,times}

\input{math_commands.tex}

\usepackage{hyperref}
\usepackage{url}
\usepackage{booktabs}
\usepackage{multirow}
\usepackage{amssymb}
\usepackage{pifont}
\usepackage{xcolor}
\usepackage{textcomp}
\usepackage{graphicx}
\usepackage[version=4]{mhchem}
\usepackage{siunitx}          
\usepackage{cleveref}    
\usepackage{wrapfig}

\title{UniLabOS: An AI-Native Operating System for Autonomous Laboratories}

\author{
\textbf{Jing Gao}$^{1,2,3,\dagger}$ \quad \textbf{Junhan Chang}$^{1,4,\dagger,*}$ \quad \textbf{Haohui Que}$^{5,6,7}$ \quad \textbf{Yanfei Xiong}$^{1}$\\
\textbf{Shixiang Zhang}$^{1}$ \quad \textbf{Xianwei Qi}$^{1}$ \quad \textbf{Zhen Liu}$^{1}$ \quad \textbf{Jun-Jie Wang}$^{1}$ \quad \textbf{Qianjun Ding}$^{1}$\\
\textbf{Xinyu Li}$^{1,8}$ \quad \textbf{Ziwei Pan}$^{1}$ \quad \textbf{Qiming Xie}$^{1}$ \quad \textbf{Zhuang Yan}$^{6}$\\
\textbf{Junchi Yan}$^{3,*}$ \quad \textbf{Linfeng Zhang}$^{1,*}$\\[0.5em]
$^{1}$DP Technology \enspace $^{2}$Zhongguancun Academy \enspace $^{3}$Shanghai Jiao Tong University\\
$^{4}$Peking University \enspace $^{5}$Shanghai Innovation Institute \enspace $^{6}$AI for Science Institute, Beijing\\
$^{7}$East China Normal University \enspace $^{8}$Beijing University of Chemical Technology\\[0.3em]
$^{\dagger}$Equal contribution \quad $^{*}$Corresponding authors: \texttt{changjh@dp.tech}, \texttt{yanjunchi@sjtu.edu.cn}, \texttt{zhanglf@dp.tech}
}

\begin{document}

\maketitle

\begin{abstract}
Autonomous laboratories promise to accelerate discovery by coupling learning algorithms with robotic experimentation, yet adoption remains limited by fragmented software that separates high-level planning from low-level execution. Here we present \textbf{UniLabOS}, an \emph{AI-native operating system} for autonomous laboratories that bridges digital decision-making and embodied experimentation through typed, stateful abstractions and transactional safeguards. UniLabOS unifies laboratory elements via an \emph{Action/Resource/Action\&Resource} (A/R/A\&R) model, represents laboratory structure with a dual-topology of logical ownership and physical connectivity, and reconciles digital state with material motion using a transactional \emph{CRUTD} protocol.

Built on a distributed edge--cloud architecture with decentralized discovery, UniLabOS enables protocol mobility across reconfigurable topologies while supporting human-in-the-loop governance. We demonstrate the system in four real-world settings—a liquid-handling workstation, a modular organic synthesis platform, a distributed electrolyte foundry, and a decentralized computation-intensive closed-loop system—showing robust orchestration across heterogeneous instruments and multi-node coordination. UniLabOS establishes a scalable foundation for agent-ready, reproducible, and provenance-aware autonomous experimentation.
\end{abstract}

\section{Introduction}
Over the past decade, laboratory automation has evolved from a throughput-enhancement tool into foundational infrastructure that is reshaping how scientific hypotheses are generated, tested, and refined. By integrating robotic platforms, online characterization, and computational planning into closed-loop workflows, a growing body of work has compressed the ``Design--Make--Test--Analyze'' (DMTA) cycle---often spanning months in traditional settings---into timeframes that are difficult to achieve with manual execution alone. This shift has catalyzed Self-Driving Laboratories (SDLs) and Materials Acceleration Platforms, which aim to systematically accelerate materials and molecular discovery~\cite{abolhasani_rise_2023,macleod_self-driving_2020}. Crucially, these systems are \emph{decision-driven}: they couple learning and optimization with execution to enable iterative hypothesis generation, validation, and refinement. Closed-loop research has already delivered measurable advances across domains, including Bayesian active learning on synchrotron beamlines for on-the-fly phase mapping and property optimization~\cite{kusne_--fly_2020}, machine-learning-guided polymer synthesis for multi-objective optimization~\cite{reis_machine-learning-guided_2021,coley_robotic_2019}, and probabilistic optimizers tailored for chemistry~\cite{roch_chemos_2018,hase_next-generation_2019}.

Several platforms have also achieved end-to-end autonomy that validates the SDL paradigm in real laboratories. ``A-Lab'' targeted 58 predicted compounds and synthesized 41 novel inorganic solids in 17 days of continuous operation, bridging computational screening and experimental realization~\cite{szymanski_autonomous_2023}. The ``Mobile Robotic Chemist'' autonomously navigated a laboratory environment and executed solid weighing and photocatalytic experiments, discovering catalyst formulations with substantially improved activity over baselines~\cite{burger_mobile_2020}. Self-driving physical vapor deposition systems have further demonstrated real-time optimization of thin-film properties via adaptive control of process parameters such as temperature and composition, mitigating batch-to-batch variability through calibration strategies~\cite{zheng_self-driving_2025,vasudevan_autonomous_2021}. Collectively, these results suggest that SDLs are moving beyond proofs of concept. Looking forward, SDLs offer value beyond throughput by generating structured, provenance-rich training data for learning systems that must reason over constraints, uncertainty, and multi-property trade-offs~\cite{kusne_--fly_2020,abolhasani_rise_2023}, and by supporting remote and distributed experimentation coordinated by cloud-based planners~\cite{hase_phoenics_2018,christensen_data-science_2021}.

Despite these frontier successes, the daily operation of most research and industrial laboratories remains constrained by a software ``reality gap'' between high-level information management and low-level experimental execution. Electronic Laboratory Notebooks (ELNs) and Laboratory Information Management Systems (LIMS) are widely deployed, but they are primarily designed for compliance, sample tracking, and retrospective archiving. As a result, they typically record \emph{what happened} after an experiment rather than serving as an execution substrate that can decide and enforce \emph{what should happen next}. In particular, they lack an explicit \emph{action layer} for dynamic device control, real-time feedback, and the execution of experimental logic. Consequently, the semantic link between physical execution and resulting data is often weak or broken, leaving laboratories without a ``central nervous system'' that can coordinate models and planners with instruments and material handling.

This gap is compounded by fragmentation in hardware and data. Modern laboratories combine instruments from multiple vendors and eras, exposing heterogeneous interfaces---from legacy RS-232 and GPIB to USB, Ethernet, and proprietary protocols. Such heterogeneity creates a prohibitive integration tax: each new device often requires bespoke driver development and fragile glue code, a barrier repeatedly identified as a bottleneck for SDL deployment at scale~\cite{stach_autonomous_2021}. Meanwhile, data silos arise because devices emit outputs in incompatible formats (e.g., CSV, XML, binary) with inconsistent metadata schemas. Critical execution context---including device configurations, calibration states, and environmental conditions---is frequently captured in unstructured notes or lost entirely, preventing the construction of traceable, end-to-end experimental graphs required for reliable learning and reproducibility.

To address orchestration challenges, several ``laboratory operating system'' prototypes have been proposed. ChemOS provides a structured middleware layer that connects AI planners to automated equipment for remote control and closed-loop optimization~\cite{roch_chemos_2018,hase_next-generation_2019}. ChemOS~2.0 treats the laboratory explicitly as an operating system and emphasizes coordination of communication, data exchange, and instruction management among modular components, demonstrating cross-device orchestration in organic laser molecule discovery~\cite{sim_chemos_2024,tom_self-driving_2024}. AlabOS focuses on solid-state synthesis logistics, introducing resource reservation and graph-based workflow models to avoid deadlocks in high-throughput campaigns~\cite{fei_alabos_2024,tom_self-driving_2024}. OCTOPUS offers an event-driven infrastructure for distributed task optimization and job parallelization via a hybrid cloud--edge event bus~\cite{yoo_octopus_2024,blaiszik_data_2019}. Despite clear progress, persistent pain points remain: (i) high integration barriers due to the lack of standardized, reusable device drivers~\cite{seifrid_autonomous_2022,coley_autonomous_2020}; (ii) limited portability beyond specific testbeds, often with insufficient encapsulation at device and data layers~\cite{blaiszik_data_2019,stein_progress_2019}; and (iii) constrained interoperability with external systems (e.g., general AI agents, enterprise ELN/LIMS) caused by tightly coupled or domain-specific APIs~\cite{coley_autonomous_2020}.

To bridge this reality gap and dismantle technological silos in laboratory automation, we propose \textbf{UniLabOS}, an \emph{AI-native} experiment operating system that serves as a missing ``kernel'' for \emph{agent-ready} autonomous laboratories. UniLabOS is inspired by the UNIX philosophy of standard interfaces and composability, but is grounded in the requirements of embodied scientific work: typed, stateful operations over physical resources with explicit feedback, feasibility checks, and transactional safeguards. Specifically, UniLabOS introduces a standardized \emph{Action}, \emph{Resource}, and \emph{Action \& Resource} (\emph{A/R/A\&R}) abstraction that treats laboratory elements---instruments, vessels, and capabilities such as heating or stirring---as first-class objects with uniform semantics. For example, a heater--stirrer is modeled as an A\&R object that both stores and acts on material, a vial is a pure Resource that only holds material, and a peristaltic pump exposes pure Actions for dosing without owning material. To model laboratory structure, UniLabOS employs a \emph{dual-topology} representation: a hierarchical \emph{resource tree} for ownership and logical containment, and a \emph{physical graph} for feasible material and data pathways. UniLabOS further extends standard Create, Read, Update, and Delete (CRUD) operations to \emph{Create}, \emph{Read}, \emph{Update}, \emph{Transfer}, and \emph{Delete} (CRUTD), where \emph{Transfer} is a first-class operation that captures spatiotemporal constraints of material movement and distinguishes physical transport from state updates. Together with a low-coupling driver model, digital-twin synchronization, and an explicit action layer, UniLabOS enables \emph{layered autonomy}: humans remain in the loop for goals, approvals, and governance, while the system enforces safe, reproducible execution and produces provenance-rich traces suitable for learning and audit.

\paragraph{Contributions.}
This work makes three main contributions.

\textbf{(1)} We introduce UniLabOS, an AI-native operating-system-style abstraction stack for autonomous laboratories, built on unified A/R/A\&R objects, a dual-topology representation (resource tree and physical graph), and a transactional CRUTD protocol for material lifecycle management.

\textbf{(2)} We design and implement a distributed edge--cloud architecture based on Robot Operating System 2 (ROS~2) and Data Distribution Service (DDS) that supports self-organizing device enrollment, a decoupled driver-as-a-service model across tens of instruments and protocols, and digital-twin synchronization for safe pre-dispatch validation.

\textbf{(3)} We demonstrate UniLabOS in four real-world case studies of increasing complexity---from a single liquid-handling workstation to a modular organic synthesis station, a distributed electrolyte foundry, and a decentralized computation-intensive control system---showing protocol mobility, topology adaptability, and robust cross-node orchestration across heterogeneous experimental setups.

\paragraph{Roadmap.}
We first present the UniLabOS architecture and communication model, then describe how standardized drivers and digital-twin synchronization enable universal hardware interoperability, followed by unified resource virtualization with transactional spatiotemporal management and decentralized orchestration. We finally demonstrate the system through four case studies of increasing complexity, and detail the underlying mechanisms in the Methods section.

\section{Results}

\subsection{UniLabOS: An AI-native distributed operating system for laboratory virtualization}
\label{subsec:unilabos-arch}

\begin{figure}[t]
    \centering
    \includegraphics[width=1\linewidth]{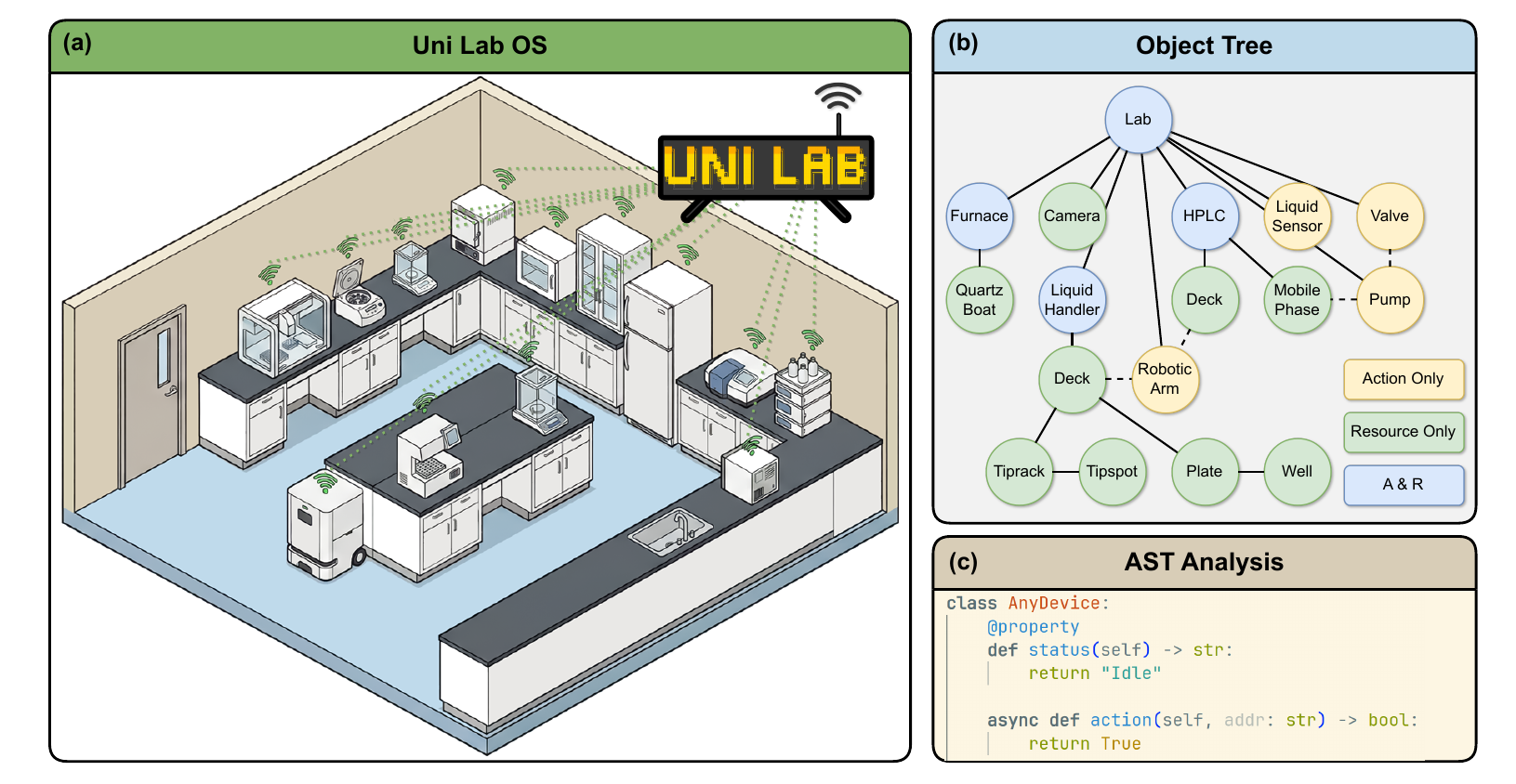}
    \caption{\textbf{UniLabOS architecture overview.}
    \textbf{(a)} A UniLabOS-managed autonomous laboratory virtualizes heterogeneous instruments---including analytical devices, liquid handlers, robotic arms, storage units, and mobile robots---as networked endpoints with standardized semantics. Devices connect to local edge nodes via wired or wireless interfaces, enabling cloud-available digital twins, AI agent integration, centralized scheduling, and unified analytics.
    \textbf{(b)} Dual-topology representation: a hierarchical \emph{resource tree} (solid lines) encodes ownership and containment (e.g., room $\rightarrow$ bench $\rightarrow$ device $\rightarrow$ container $\rightarrow$ sample) for resource query and access control; a \emph{physical graph} (dashed lines) captures feasible material pathways (fluidic tubing, robotic reachability) for transfer routing and workflow orchestration.
    \textbf{(c)} Dynamic driver registration via AST inspection: UniLabOS parses driver source code to extract method signatures and type hints, and automatically registers capabilities into the resource tree without process restart.}
    \label{fig:lab_overview}
\end{figure}

To bridge the operational gap between high-level AI planning and low-level physical execution, we developed \textbf{UniLabOS}, a distributed operating system designed to manage diverse laboratory environments---ranging from single workstations to distributed, multi-site facilities (Fig.~\ref{fig:lab_overview}a). UniLabOS virtualizes the physical laboratory into a programmable substrate, centrally managing both the hardware environment and experimental workflows to ensure reliable execution across heterogeneous networking topologies.

To achieve this virtualization, UniLabOS integrates three core technical pillars. First, it models laboratory structure using a \textbf{dual-topology} representation (logical resource tree and physical graph; Fig.~\ref{fig:lab_overview}b) and unifies heterogeneous hardware under an \textbf{A/R/A\&R} abstraction (detailed in \S\ref{subsec:resource-management}). Second, it establishes a robust, decentralized communication backbone via \textbf{ROS~2/DDS}, enabling reliable peer-to-peer messaging and self-organizing discovery (detailed in \S\ref{subsec:iot-arch}). Third, it lowers integration barriers through \textbf{Abstract Syntax Tree (AST)-based} driver analysis, which allows low-coupling code writing and dynamic capability registration (detailed in \S\ref{subsec:hardware-interop-dt}).

The reference implementation of UniLabOS\footnote{\url{https://github.com/deepmodeling/Uni-Lab-OS}} is released as open-source software and focuses on the edge and device-integration layer, enabling laboratories and instrument vendors to deploy, inspect, and extend the on-premises control stack. Bohrium\footnote{\url{https://www.bohrium.com/}} serves as a unified infrastructure for ``reading, computing, and experiment'' across simulations and physical experiments, and provides a managed cloud environment for cross-site coordination, long-term data retention, and collaborative monitoring. In this arrangement, UniLabOS provides an open and extensible kernel for hardware interoperability, while \textbf{Uni-Lab cloud services running on Bohrium} provides cloud-side operational capabilities (e.g., high-availability hosting, security hardening, and access control) for deployments that require remote coordination.

\subsection{Universal hardware interoperability via unified device abstraction}
\label{subsec:hardware-interop-dt}

To manage the complexity of modern laboratories, UniLabOS implements a decoupled \emph{driver-as-a-service} model that separates high-level intent from low-level device protocols and supports integration across communication standards including Modbus, PLC, ROS~2, OPC Unified Architecture (OPC~UA), and TCP/IP. Each device capability is exposed through standardized atomic interfaces, utilizing typed getter and setter methods (e.g., for temperature control or stirring speed). This abstraction effectively smooths over hardware heterogeneity, allowing agents and applications to interact with diverse instruments—whether a legacy serial-connected heater or a modern Ethernet-connected robot—through the same API interface.

UniLabOS categorizes hardware resources into six functional classes: \textbf{sensors} (e.g., pH meters, thermocouples, cameras); \textbf{connectors} (e.g., pumps, valves); \textbf{material processing} units (e.g., liquid handlers, powder dispensers, muffle furnaces); \textbf{characterization} instruments (e.g., HPLC, GC-MS); \textbf{logistics} systems including robotic arms and automated guided vehicles (AGVs); and \textbf{virtual devices} such as computing nodes, simulators, and encapsulated workstation systems, covering key stages of the DMTA loop. UniLabOS also acts as an abstraction layer that smooths over interface heterogeneity. By abstracting common functionalities within each category, the system implements a "single interface, multiple implementations" pattern. This standardization ensures that diverse instruments sharing the same role (e.g., different brands of liquid handlers) expose identical control semantics, allowing upper-layer applications to interact with hardware through unified, protocol-agnostic interfaces. We provide a practical demonstration of this abstraction using a liquid handling workstation in Section~\ref{subsec:case1}.

Moreover, by mapping standardized device primitives onto ROS~2 communication patterns (Methods), UniLabOS bridges laboratory automation with the broader robotics ecosystem and enables bidirectional \emph{digital-twin synchronization}. Telemetry streams continuously update the virtual state of devices and resources. Conversely, the digital twin supports pre-dispatch validation: before executing a command sequence, UniLabOS can perform a look-ahead check in a simulated environment (e.g., RViz/MoveIt) to detect collisions, kinematic infeasibility, or protocol violations that may not be captured by purely logical constraints. This reduces equipment risk and improves safety for autonomous exploration and remote operation.

\begin{figure}[t]
    \centering
    \includegraphics[width=1\linewidth]{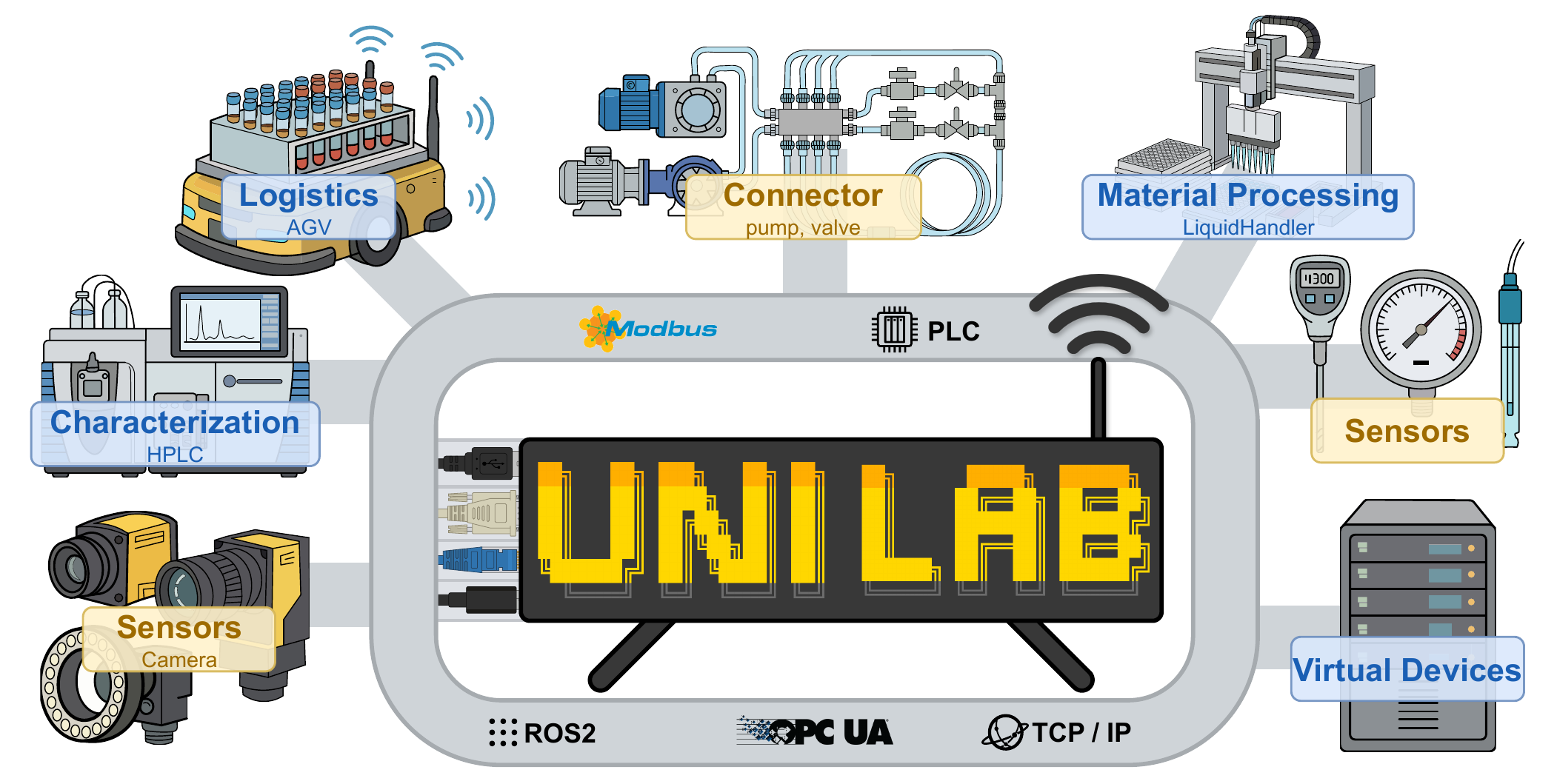}
    \caption{\textbf{Universal hardware interoperability in UniLabOS.} UniLabOS provides a unified integration layer that connects heterogeneous laboratory instruments---including fluidic systems, liquid handlers, analytical devices (e.g., HPLC, mass spectrometers), robotic arms, AGVs, sensors (pressure, pH), and data servers---through diverse communication protocols such as Modbus, PLC, ROS~2, OPC~UA, and TCP/IP. This protocol-agnostic abstraction enables seamless device enrollment and interoperability across vendor ecosystems.}
    \label{fig:devices_std}
\end{figure}

\subsection{Unified resource virtualization and transactional spatiotemporal management}
\label{subsec:resource-management}

Achieving reliable autonomy in chemical laboratories requires unified management of heterogeneous entities. As an operating system for the laboratory, UniLabOS is tasked with managing the complete experimental context: all physical resources, active devices, and the execution flows that connect them. This scope encompasses the full diversity of a classical laboratory environment (Fig.~\ref{fig:lab_overview}a), requiring a unified control plane that can represent everything from reagents and vials to complex instruments and multi-step protocols. 

To achieve this universal representation, UniLabOS introduces two complementary abstraction mechanisms. First, it classifies all laboratory entities into three fundamental types: \emph{Resources} (R) for passive material entities (e.g., vials), \emph{Actions} (A) for pure executable capabilities (e.g., pumps), and \emph{Action \& Resource} (A\&R) composites for hardware that both stores and acts on materials (e.g., liquid handler). Second, it models the structural relationships between these entities using a \textbf{dual-topology} approach. A \textbf{logical resource tree} defines ownership and containment hierarchies (e.g., Device $\subset$ Bench $\subset$ Room), while a \textbf{physical graph} captures functional connectivity, such as fluidic tubing or robotic reachability. This dual representation uniquely characterizes both the static organization and the dynamic capabilities of the laboratory—a feature critical for adaptive workflows like modular organic synthesis (showed in \S\ref{subsec:case2}).

Mirroring the flow of experiments is the flow of materials and state updates. To formalize these transitions, UniLabOS extends the standard database CRUD model to \textbf{CRUTD} by elevating \textbf{Transfer} to a first-class operation (formal definitions in Methods). While CRUD manage digital state and properties, \emph{Transfer} explicitly handles the spatiotemporal movement of materials (e.g., liquid dispensing, vial transport). By treating Transfer as an atomic transaction that couples ownership updates with physical actuation verification, CRUTD ensures that the digital record remains synchronized with physical reality, preventing state drift and enabling precise provenance tracking across complex experimental workflows.

\begin{figure}[t]
    \centering
    \includegraphics[width=0.9\linewidth]{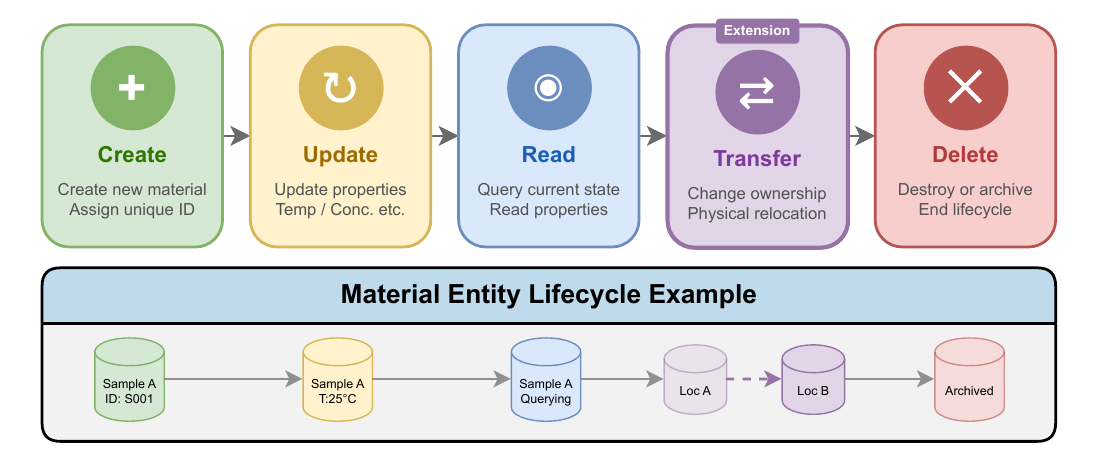}
    \caption{\textbf{The CRUTD operation set for material lifecycle management.} \textbf{Create} introduces new entities; \textbf{Update} modifies properties without changing ownership; \textbf{Read} queries state; \textbf{Transfer} relocates materials across spatial or logical boundaries and explicitly captures spatiotemporal constraints; \textbf{Delete} removes entities from active workflows. Representing laboratory workflows as CRUTD sequences yields a unified, provenance-aware execution record.}
    \label{fig:crutd_show}
\end{figure}

\subsection{Scalable orchestration via decentralized self-organizing infrastructure}
\label{subsec:iot-arch}

Autonomous laboratories require both reliable local control and scalable orchestration across devices and sites. UniLabOS separates the intranet (mapped to the Local Area Network, LAN) from the extranet (mapped to the Wide Area Network, WAN): edge nodes handle safety interlocks, device actuation, and time-sensitive control within the intranet, while cloud services support asynchronous scheduling, global data aggregation, and cross-site coordination within the extranet. This boundary isolates physical safety from network variability and supports \emph{human-in-the-loop} oversight through monitoring, approval, and audit.

\begin{figure}[t]
    \centering
    \includegraphics[width=1\linewidth]{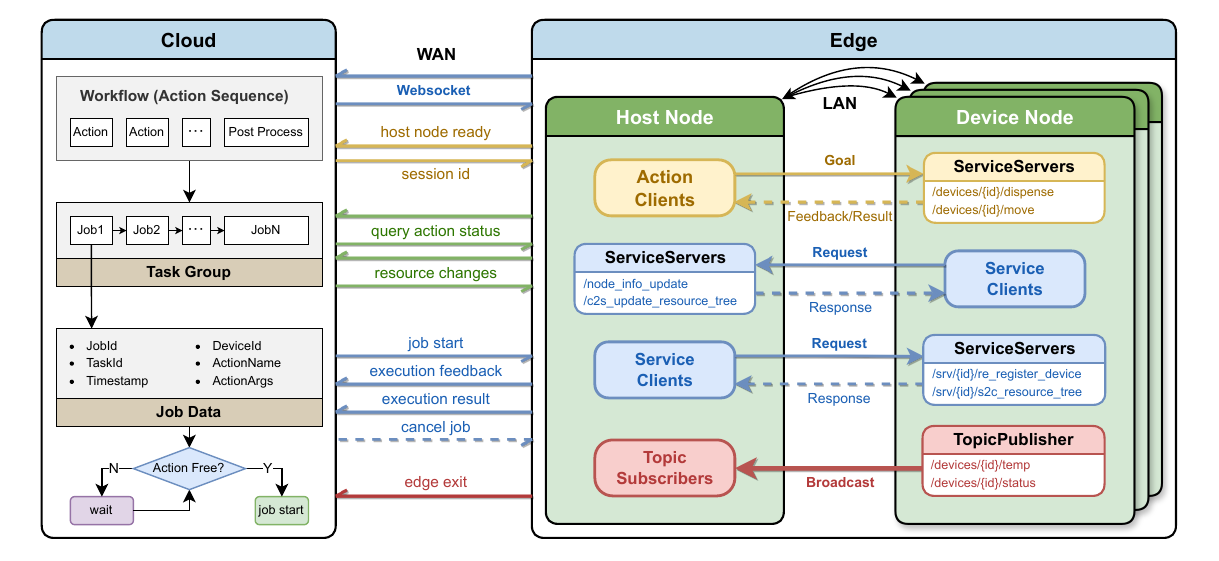}
    \caption{\textbf{Host--device communication architecture in UniLabOS.} A host node bridges the cloud/external domain and manages device nodes via ROS~2/DDS. Four channels are used: (1) \textbf{Networking} (\emph{Services}) for discovery/registration; (2) \textbf{Material} (\emph{Services}) for resource-tree synchronization with transactional checks; (3) \textbf{Action} (\emph{ROS~2 Actions}) for long-running operations with feedback and cancellation; (4) \textbf{Status} (\emph{Topics}) for telemetry broadcast (e.g., temperature, position).}
    \label{fig:ros2_network}
\end{figure}

UniLabOS employs \textbf{ROS~2 over DDS} rather than a centralized broker. DDS provides decentralized discovery and peer-to-peer communication, improving robustness under partial node failures. Notably, the peer-to-peer architecture significantly benefits scenarios where sensing and computation are distributed across devices: nodes can exchange data directly without routing through a central host, reducing latency and bandwidth bottlenecks. Building on this design, UniLabOS supports \textbf{self-organizing enrollment}: when a new device node joins the intranet, it advertises its capabilities and registers with the host via a discovery-and-handshake protocol (Fig.~\ref{fig:ros2_network}), enabling topology changes and incremental scaling with minimal reconfiguration.

Task orchestration follows an \emph{intent-to-execution} pipeline. The cloud orchestrates sequences of device actions on laboratory materials, which the scheduler converts into jobs containing all essential metadata. An entire workflow is encapsulated as a task group comprising all pending and completed jobs. After transmission over the WAN, the local host node, acting as a gateway, verifies that devices are not occupied before dispatch. Actual execution is delegated to edge devices on the local LAN and requires no further cloud participation. Thus, even during network failures, the host node sustains operations, executing queued tasks and coordinating peer-to-peer device messaging autonomously.

\subsection{Case Study 1: High-fidelity and flexible liquid handling on a single workstation}
\label{subsec:case1}

\begin{figure}[t]
    \centering
    \includegraphics[width=1\linewidth]{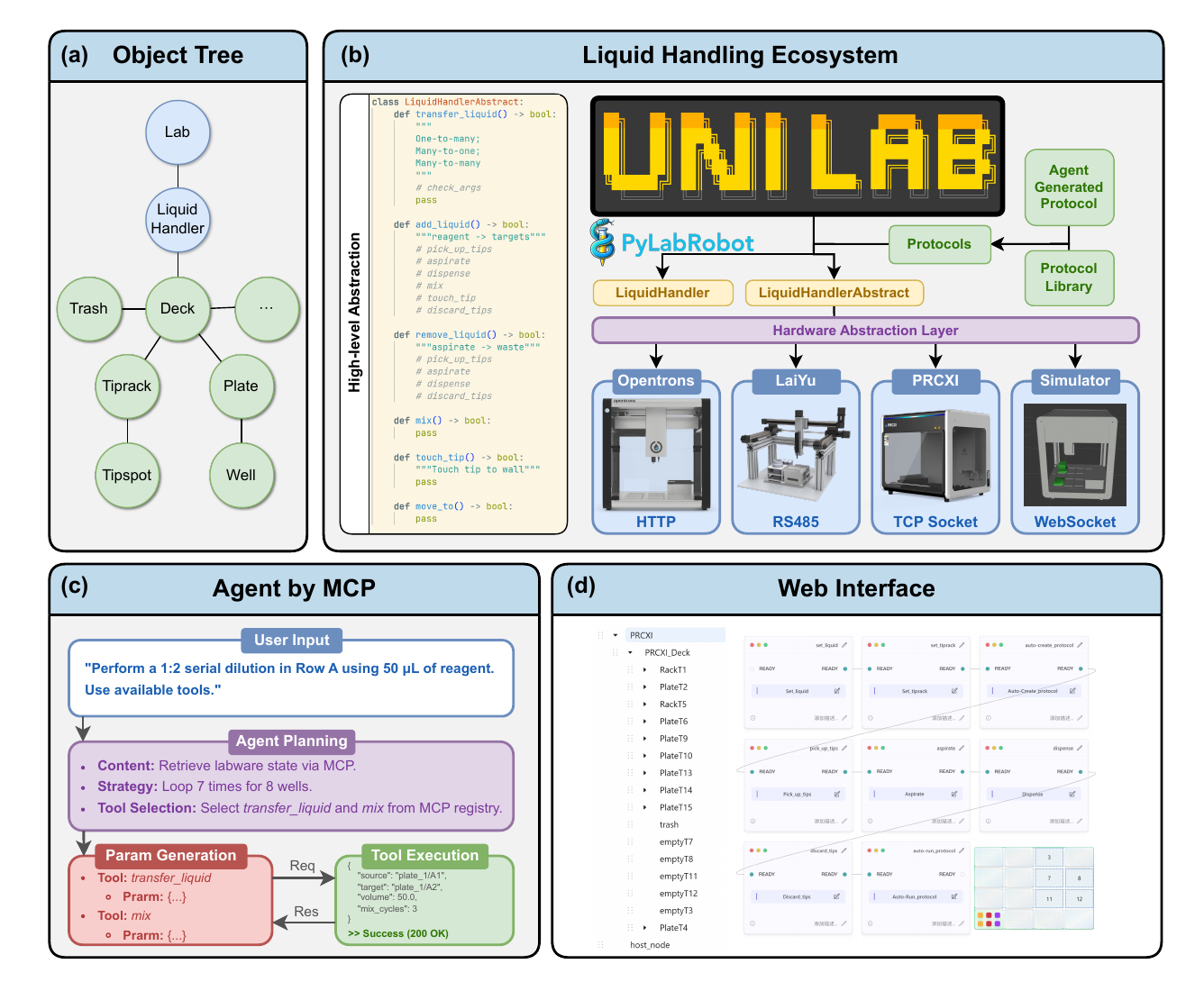}
    \caption{\textbf{Liquid handling virtualization in UniLabOS.}
    \textbf{(a)} Object Tree: A hierarchical representation of the liquid-handling deck, mapping containment from the deck down to plates and individual wells.
    \textbf{(b)} Liquid Handling Ecosystem: The layered architecture compiles protocols through the \texttt{LiquidHandlerAbstract} (LHA) interface into vendor-specific backends (e.g., PRCXI, Laiyu, simulator), ensuring hardware-agnostic execution.
    \textbf{(c)} Agent by MCP: An LLM agent uses the Model Context Protocol (MCP) to query labware states and generate experimental plans.
    \textbf{(d)} Web Interface: The generated plans are executed and visualized within the UniLab platform interface, which displays real-time workflow status and resource tracking.}
    \label{fig:lh_arch}
\end{figure}

We demonstrate the versatility of UniLabOS on liquid-handling workstations—representing a \emph{single-host, single-device} configuration—through a three-layered architecture (Fig.~\ref{fig:lh_arch}). Liquid handling workstations (Fig.~\ref{fig:lh_arch}a) are typically composed of a deck holding various labware resources, a gantry system (3-axis motors), and fluidic controllers. UniLabOS introduces a unified Hardware Abstraction Layer that decouples logical intent from physical implementation, enabling seamless compatibility across diverse brands and simulation environments. This abstraction layer manages the decomposition of high-level commands into device-specific driver signals for material manipulation. In practice, our implementation supports a wide range of backend interfaces, including HTTP, RS485, and TCP Socket communication for liquid handlers, and WebSocket-based integration for RViz visualization and simulation. Moreover, by encoding liquid transfer and state updates as standard CRUTD transactions, we ensure precise and traceable material provenance throughout the workflow. 

Building upon basic primitives (e.g., from PyLabRobot), we designed the \texttt{LiquidHandlerAbstract} interface to expose topology-aware unit operations: \texttt{transfer}, \texttt{add}, \texttt{remove}, and \texttt{mix}. These atomic actions significantly reduce the complexity and redundancy of protocol authoring compared to low-level aspirate--dispense sequences. This semantic compression allows Large Language Models (LLMs) to more easily generate correct workflows with high reliability. Furthermore, these unit operations facilitate rigorous pre-execution validation: the system leverages both a physics-based liquid handling simulator and RViz-based collision detection to perform logical checks (e.g., volume tracking) and kinematic verification (e.g., safe loading/unloading paths), ensuring experimental safety and enabling high-throughput virtual screening.

We also enable natural language-driven experimentation through an agentic workflow (Fig.~\ref{fig:lh_arch}c). The process follows a structured pipeline: User Input $\rightarrow$ Agent Planning $\rightarrow$ Tool Selection $\rightarrow$ Parameter Generation $\rightarrow$ Tool Execution. Using the Model Context Protocol (MCP), an LLM agent queries the live labware state, plans the experiment, and compiles the intent into executable driver actions. This pipeline culminates in the UniLab platform interface (Fig.~\ref{fig:lh_arch}d), which visualizes real-time workflow status and resource tracking, bridging the gap between abstract scientific intent and precise hardware execution.

\subsection{Case Study 2: Adaptive topology for organic synthesis}
\label{subsec:case2}

We next evaluated UniLabOS in a \emph{single-host, multi-device} organic synthesis setting. In modern organic chemistry, compound preparation still necessitates substantial hands-on time from researchers for repetitive manual operations; automating these procedures can meaningfully liberate time for higher-value tasks such as reaction design and analysis~\cite{steiner_organic_2019}.

\begin{figure}[t]
    \centering
    \includegraphics[width=1\linewidth]{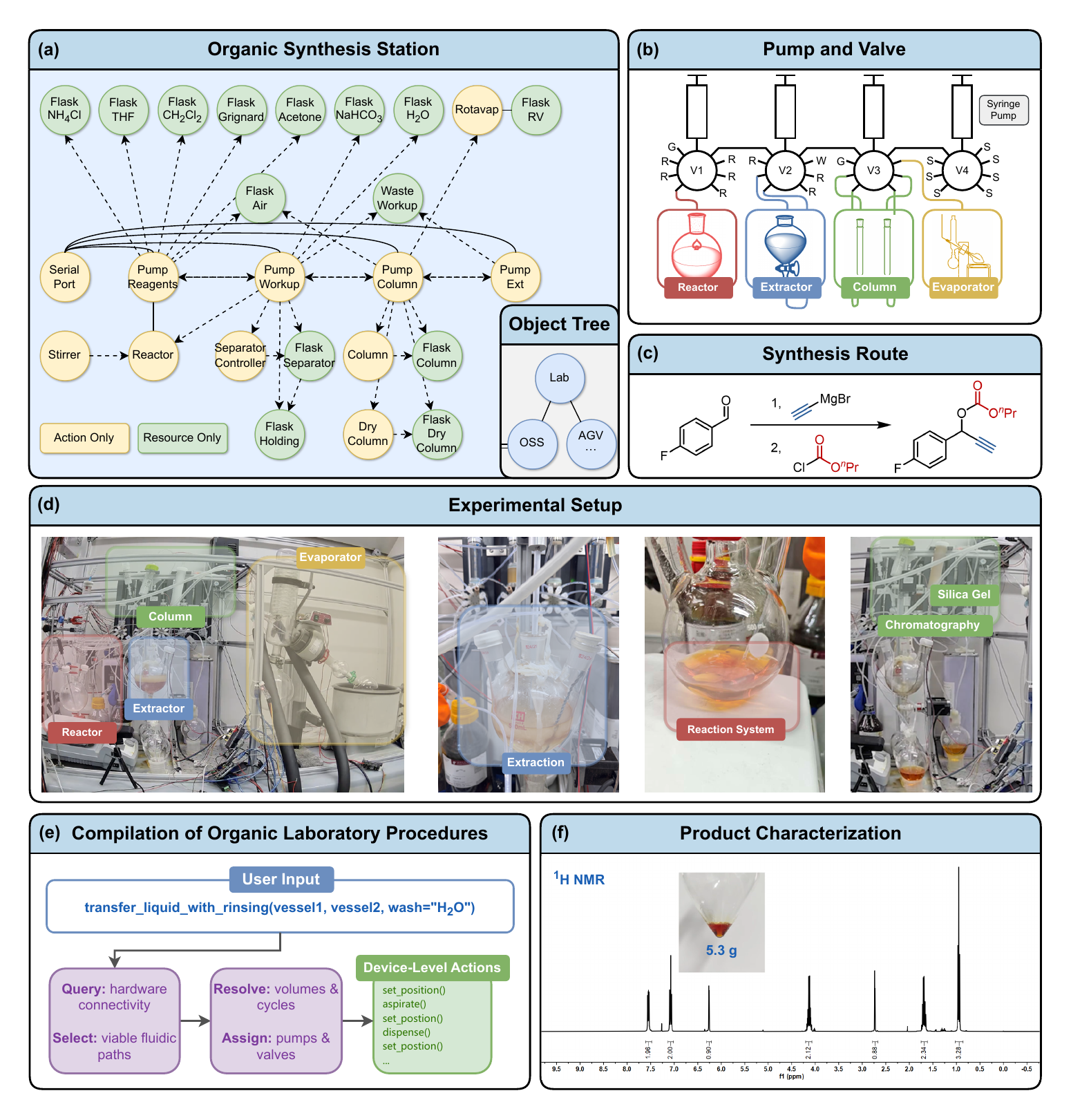}
    \caption{\textbf{Modular organic synthesis workstation.}
    \textbf{(a)} Physical-graph representation of the laboratory hardware. Nodes represent devices (yellow) and resources (green); dashed arrows indicate feasible material pathways. An object tree (inset) shows the hierarchical relationship.
    \textbf{(b)} Schematic of the multi-port valve and syringe pump assembly, illustrating connectivity between four valves (V1–V4) and ports for reagents (R), gas (G), waste (W), and solvents (S), routed to the reactor, extractor, column, and evaporator.
    \textbf{(c)} Synthesis route.
    \textbf{(d)} Experimental setup photographs displaying the integrated hardware components including the reaction system, extraction unit, and chromatography station.
    \textbf{(e)} Software architecture for compiling organic laboratory procedures. High-level user input is translated into device-level actions via hardware connectivity queries, path selection, volume/cycle resolution, and pump/valve assignment.
    \textbf{(f)} Product characterization. $^{1}$H-NMR spectrum of the product (5.3\,g) synthesized by the workstation.}
    \label{fig:organic_real}
\end{figure}

To this end, we developed an automated organic synthesis workstation for substrate preparation. The workstation integrates temperature-controlled stirring and automated reagent delivery, along with downstream modules for automated liquid--liquid extraction, silica flash chromatography, and rotary evaporation (Fig.~\ref{fig:organic_real}b).

UniLabOS's \textbf{dual-topology} model---capturing both a logical resource tree (hierarchical ownership) and a physical graph (functional connectivity)---is particularly advantageous in this scenario. The logical resource tree facilitates efficient device scheduling and structured data organization by grouping sensors and actuators under logical parent nodes. Simultaneously, the physical graph (Fig.~\ref{fig:organic_real}a) enables the compilation of abstract protocols (e.g., XDL~\cite{rauschen_universal_2024}) into valid transfer paths. This allows the same synthesis recipe to be seamlessly migrated across different physical setups by simply updating the connectivity graph, demonstrating robust decoupling between experimental logic and physical realization.

We executed a representative two-step reaction starting from an aldehyde: nucleophilic addition of a Grignard reagent affords the corresponding alcohol intermediate, which is subsequently esterified with an acyl chloride to yield the target propargyl ester (Fig.~\ref{fig:organic_real}c). After reaction completion, the mixture was quenched and subjected to multiple aqueous washes to remove inorganic salts. The organic phase was dried, followed by a brief silica flush with dichloromethane (DCM) and solvent removal by rotary evaporation to afford the final product (Fig.~\ref{fig:organic_real}d). Using this automated pipeline, we obtained 5.3\,g of product, meeting preparative-scale requirements, with $^{1}$H NMR confirming spectroscopic purity (Fig.~\ref{fig:organic_real}f).

\subsection{Case Study 3: Cross-node coordination in a distributed electrolyte foundry} \label{subsec:case3}

To evaluate scalability across physically isolated domains, we deployed UniLabOS in a multi-host facility for high-throughput electrolyte research, comprising three distinct workstations: an inert-atmosphere formulation glovebox, a coin-cell assembly station, and a battery testing array. Virtualizing this discontinuous physical space requires mapping the movement of materials between isolated environments into a single logical resource tree. A global coordination server manages the overarching workflow, while local host nodes independently control peripherals—including robotic arms and AGVs—forming a hierarchical architecture that centralizes data management while distributing execution.

A critical capability in this distributed setting is maintaining a continuous chain of custody as materials undergo transformation. The workflow begins with raw lithium salts, solvents and functional additives, tracked as distinct resources, being combined into a unified formulation container. UniLabOS records this genesis via CRUTD transactions that link raw material batches to the new mixture. As the electrolyte is transferred to the assembly station and injected into cells, the system creates nested cell assembly objects that logically encapsulate electrode layers and the specific electrolyte source. This object-oriented provenance ensures that downstream electrochemical data is automatically linked back to the precise formulation and assembly conditions, establishing a comprehensive material lineage without manual alignment.

We further validated the system's resilience to network instability, a key requirement for multi-site operations. In a stress test where the assembly host was disconnected, the global scheduler utilized decoupled instruction queues to manage the fault. Upon detecting heartbeat loss, the scheduler isolated the command queue for the assembly node, pausing its dependent tasks, while the independent queue for the formulation station continued to process tasks uninterrupted. Upon reconnection, a handshake process synchronized the resource states between the local and global trees. This mechanism confirms UniLabOS's ability to sustain robust orchestration through granular task isolation, ensuring data integrity even in imperfect network environments.

\begin{figure}[t]
    \centering
    \includegraphics[width=1\linewidth]{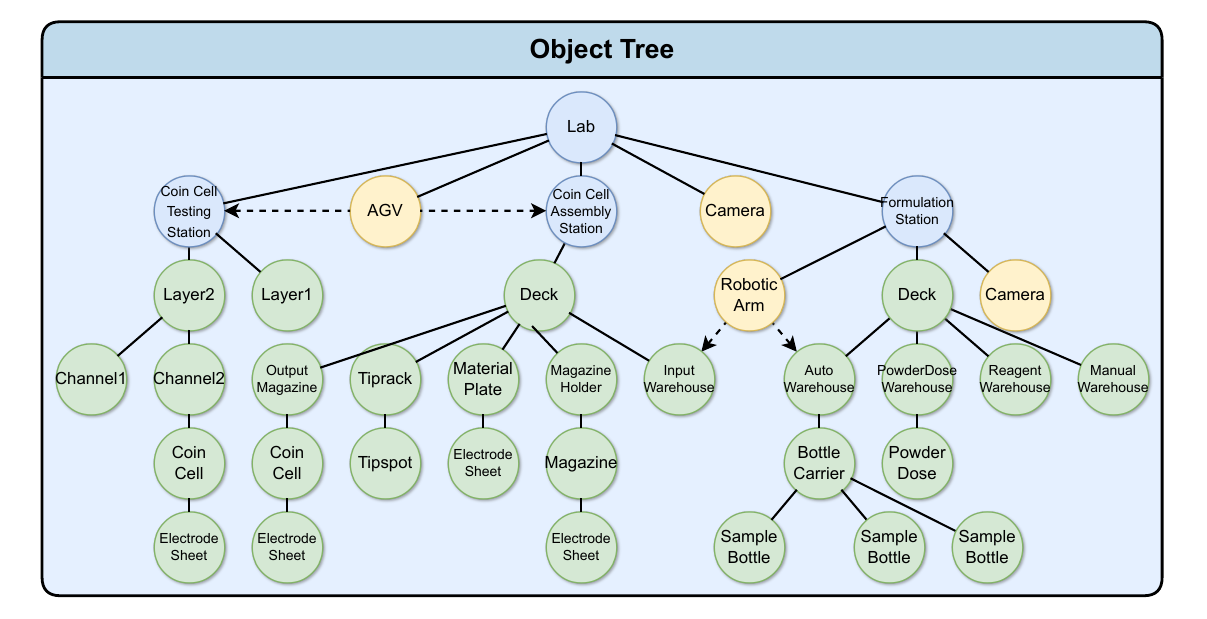}
    \caption{\textbf{Distributed electrolyte foundry resource topology.} Three hosts (glovebox formulation, coin-cell assembly, and testing) are coordinated via UniLabOS.}
    \label{fig:elec_resource}
\end{figure}

\subsection{Case Study 4: Decentralized Orchestration for Computation-Intensive Closed-Loops}
\label{subsec:case4}

\begin{figure}[t]
    \centering
    \includegraphics[width=1\linewidth]{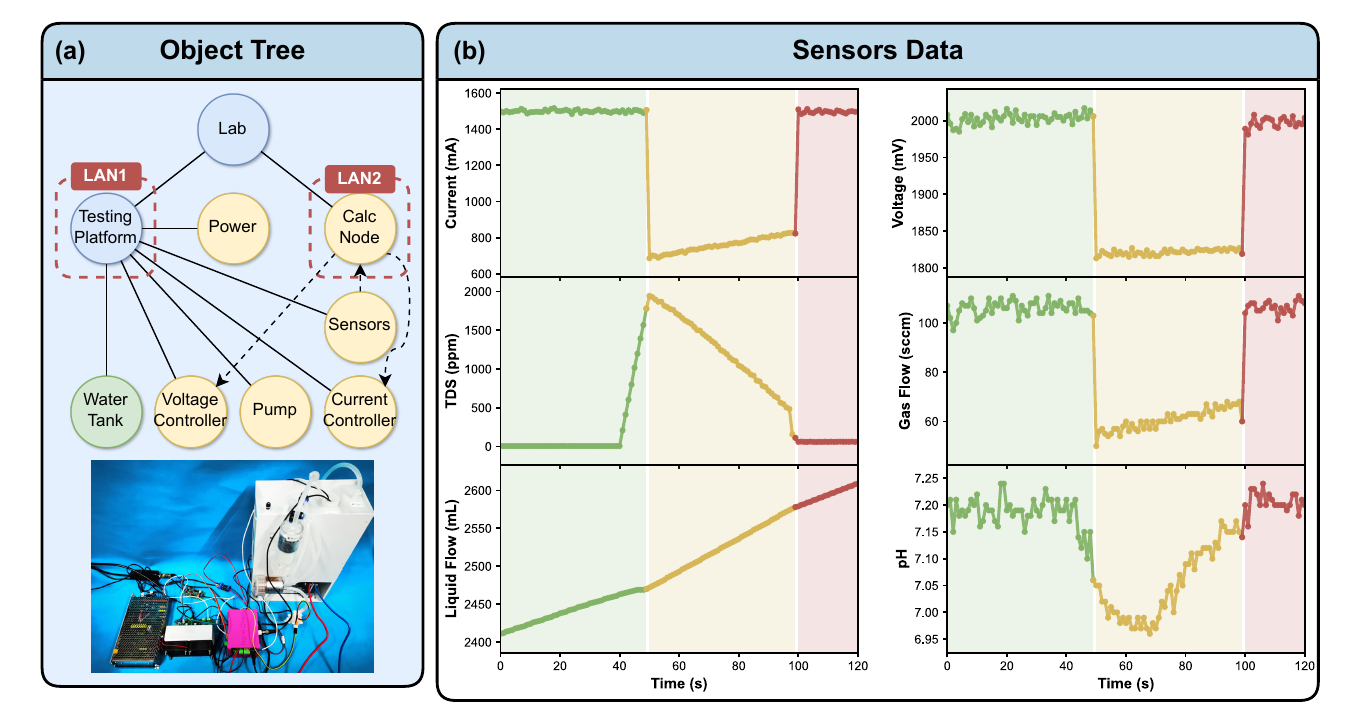}
    \caption{\textbf{Decentralized orchestration for computation-intensive closed-loops.} (a) Object tree showing the logical separation of data acquisition and compute nodes. (b) Experimental validation: real-time TDS monitoring triggers autonomous mode switching (CC to CV) via peer-to-peer communication.}
    \label{fig:case4}
\end{figure}

To demonstrate UniLabOS's resilience in high-demand, distributed environments (Fig.~\ref{fig:case4}), we implemented a decoupled architecture where data acquisition nodes are physically and logically separated from computational nodes. In conventional laboratory frameworks, real-time telemetry is typically routed through a central host to cloud-based analysis nodes. Such centralized topologies are inherently vulnerable to WAN instability and host-level failures, which can terminate time-sensitive control loops. UniLabOS utilizes its ROS~2/DDS-based decentralized backbone to enable direct peer-to-peer (P2P) communication between sensors and high-performance computing (HPC) nodes. This allows the system to offload intensive time-series modeling to local clusters, bypassing the host node to minimize latency and bandwidth consumption while ensuring control continuity during external network outages.

The experimental verification involves an alkaline electrolyzed water system, where real-time total dissolved solids (TDS) determine the catalyst protection strategy. Prevent the catalyst from being severely damaged and reducing its service life due to water quality under high current and voltage conditions. The system initially operated in a preset constant current (CC) mode of 1500\,mA. When TDS surged to 1965\,ppm was detected. The remote computing node processes the input through a direct P2P data stream and issues instructions to the controller to switch the system operation mode to a constant voltage (CV) mode of 1.82\,V. The current drops to $\sim$0.69\,A and the gas production flow rate decreases synchronously, while the circulating water flow rate increases to promote water quality recovery. When TDS falls back and stabilizes, the timing control strategy automatically switches back to the preset constant current mode, and the normal gas production capacity is restored accordingly.

\section{Discussion}

\paragraph{From brittle automation to layered autonomy.} 
The field of laboratory automation has historically been stymied by a ``Tower of Babel'' problem: while individual instruments are highly capable, their mutually incompatible interfaces force researchers to serve as ad hoc system integrators. Our results suggest that an operating-system-centric philosophy---defined by the strict decoupling of scientific intent from physical implementation---is a foundational requirement for scaling SDLs. UniLabOS facilitates a paradigm of layered autonomy: human researchers define high-level objectives and governance policies, while the OS kernel manages the low-level complexities via the A/R/A\&R abstraction. This separation allows the system to handle hardware heterogeneity and resource conflict resolution, transforming the laboratory from a collection of scripts into a deterministic, managed runtime environment.

\paragraph{Semantic mobility and topological invariance.} 
By resolving semantic commands against a live dual-topology representation---comprising a logical resource tree and a physical connectivity graph---UniLabOS achieves protocol mobility. Unlike rigid scripts reliant on hard-coded coordinates, our approach allows experimental procedures to be treated as ``portable code,'' capable of adapting to different physical layouts as demonstrated in the modular organic synthesis case. This topological invariance addresses the reproducibility crisis by making environmental dependencies explicit. However, the limits of this mobility are defined by the ``hardware reality gap''---differences in instrument precision and calibration---necessitating the future development of shared semantic contracts and automated calibration routines to ensure that a protocol's execution is as standardizable as its code.

\paragraph{Bridging the symbolic--physical gap: The AI-native safety kernel.}
A critical bottleneck in AI-driven discovery is the ``symbolic gap'' between an LLM agent's high-level reasoning and the robot's embodied actuation. UniLabOS bridges this gap by providing semantic grounding through typed, stateful APIs. As shown in our liquid-handling agent, the OS acts as a safety kernel for non-deterministic agents. By enforcing resource locking, type checking, and pre-dispatch feasibility simulation, the system ensures that even if an agent's plan is logically flawed or ``hallucinated,'' the physical execution remains within the bounds of safety. Future work must focus on quantifying agent reliability and developing automated recovery strategies for exceptions that occur at the physical interface.

\paragraph{CRUTD: A transactional foundation for holistic provenance.} 
Traditional data management often captures the result but loses the context. The CRUTD protocol reconceptualizes material handling as a transactional operation, ensuring that the digital state and physical reality remain in strict synchronization. This creates a high-fidelity provenance graph that links every measurement to its specific spatiotemporal lineage---the exact pump, vial, and environmental conditions at the moment of transfer. Such data density is essential for the next generation of machine learning models that seek to learn the underlying causal structures of chemical processes (e.g., in our electrolyte foundry), rather than mere correlations between inputs and outputs.

\paragraph{Decentralized resilience and the De-siloing of ecosystems.} 
Moving beyond centralized control, UniLabOS validates a decentralized architecture based on ROS~2 and DDS. As evidenced by the fault-tolerant electrolysis experiment, this peer-to-peer capability allows edge nodes to maintain critical control loops even when connectivity to the central host is severed. This architecture not only enhances resilience but also supports a ``Lab-as-a-Service'' (LaaS) model by turning fragmented equipment into networked, auditable assets. By balancing an open-source kernel for device integration with managed cloud infrastructure (e.g., Bohrium) for multi-site orchestration, UniLabOS provides a scalable path to de-silo laboratory ecosystems without compromising local safety or operational continuity.

\paragraph{Adaptive scheduling and enhanced feedback for AI agents.} While UniLabOS successfully unifies diverse hardware within local networks via the A/R/A\&R model, supporting complex AI agents and multi-user concurrency requires evolving beyond simple execution. Future schedulers must support policy-driven orchestration to dynamically balance competing objectives, such as handling "user-priority" overrides against "throughput-optimized" background tasks, or enabling AI agents to preemptively allocate resources for exploration. Furthermore, to empower true agent autonomy, the system must extend beyond action primitives to provide multimodal observation spaces—integrating vision, spectroscopy, and internal device states—alongside standardized error recovery mechanisms. These enhancements will enable agents to not only execute protocols but also autonomously diagnose and recover from physical anomalies.

\paragraph{Towards asynchronous coordination in a global Laboratory-as-a-Service network.} As we transition from local, real-time control to a distributed "Internet of Laboratories," the challenge shifts to asynchronous coordination across geographically dispersed (inter-city) nodes. Realizing the LaaS vision entails workflows that span physical boundaries, requiring UniLabOS to extend its scope from benchtop management to the lifecycle tracking of material states across logistics chains (e.g., stability monitoring during transport). Simultaneously, global resource orchestration demands robust frameworks for cross-institutional identity management and multi-tenancy isolation. By addressing these needs, UniLabOS serves as the foundational kernel for transforming local automation into a secure, scalable, and planetary-scale scientific collaboration.

\section{Conclusion}

In conclusion, UniLabOS establishes a foundational operating system that bridges the operational gap between high-level computational planning and low-level experimental execution. By unifying heterogeneous hardware through the \textbf{A/R/A\&R abstraction} and managing laboratory structure via a \textbf{dual-topology model}, we effectively decouple scientific intent from physical complexity. This virtualization does more than simplify automation; it transforms the laboratory into a deterministic, programmable substrate, acting as a \textbf{safety kernel} that allows embodied AI agents to interact reliably with the physical world.

Furthermore, our implementation of the transactional \textbf{CRUTD protocol} and the decentralized edge--cloud architecture ensures that this autonomy is both data-rich and resilient. By enforcing strict synchronization between the digital twin and physical reality, UniLabOS guarantees the high-fidelity provenance required for data-driven research and machine learning applications. Looking ahead, we envision UniLabOS as the enabler of a global ``Internet of Laboratories,'' where standardized drivers and protocol mobility foster a collaborative ``Lab-as-a-Service'' ecosystem. Ultimately, this work provides the robust infrastructure necessary to scale Self-Driving Laboratories from isolated testbeds into a distributed, agent-ready scientific network.

\section{Methods}

\subsection{Unified resource data schema and state modeling}
\label{subsec:method-resourcedict}

To provide a uniform digital representation for heterogeneous laboratory entities, UniLabOS defines a \texttt{ResourceDict} schema as the canonical model for all \emph{Resources} (R) and \emph{Action \& Resource} (A\&R) composites. Each entity is assigned an immutable Universally Unique Identifier (UUID) as a stable primary key for lifecycle tracking independent of physical location or mutable attributes. Logical containment is represented through explicit \texttt{parent\_uuid} references, forming a recursive hierarchy (e.g., \textit{Well} within \textit{Plate} on a \textit{Deck}) that supports query, scheduling scope, and access control.

To support physical execution and digital-twin fidelity, \texttt{ResourceDict} optionally includes a spatial representation comprising a 6-DOF pose and geometric dimensions under a shared coordinate frame, enabling collision checking, reachability reasoning, and 3D rendering. The schema separates relatively stable specifications from runtime state by partitioning fields into three namespaces:
(i) \texttt{config} for static specifications and constraints (e.g., max volume, material compatibility, calibrated ranges);
(ii) \texttt{data} for mutable runtime state (e.g., current volume, temperature, occupancy);
(iii) \texttt{extra} for contextual metadata and provenance annotations (e.g., batch identifiers, run tags, operator notes).
This separation reduces redundancy and allows the same object instance to serve persistence, UI visualization, and motion-planning needs without duplicating representations.

\subsection{The CRUTD protocol and transactional state reconciliation}
\label{subsec:method-crutd}

\begin{figure}[t]
    \centering
    \includegraphics[width=0.85\linewidth]{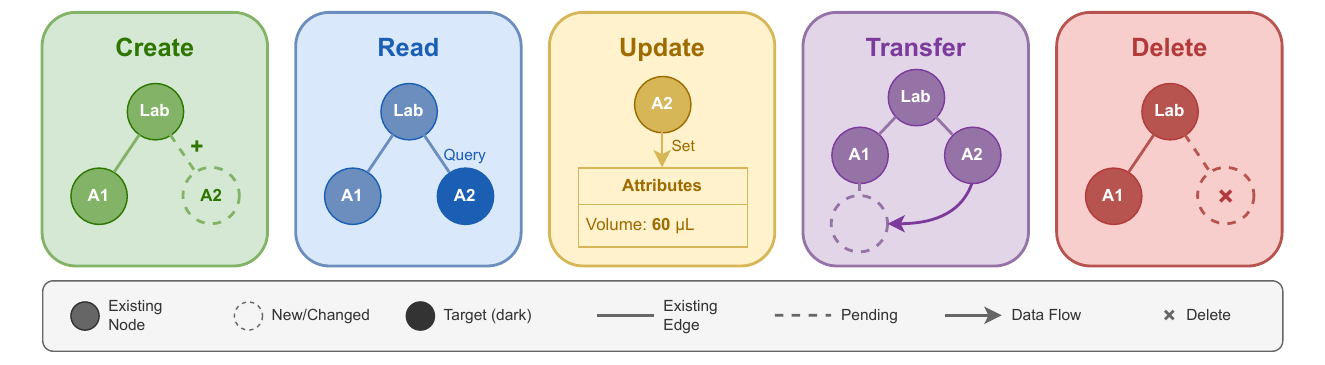}
    \caption{\textbf{Formal definition of CRUTD operations.} Each primitive (\emph{Create}, \emph{Update}, \emph{Read}, \emph{Transfer}, \emph{Delete}) is specified with pre-conditions, post-conditions, and its mapping to physical actions and feedback, enabling transactional consistency between the digital model and physical execution.}
    \label{fig:crutd}
\end{figure}

UniLabOS maps laboratory state changes onto the CRUTD command set (\emph{Create}, \emph{Read}, \emph{Update}, \emph{Transfer}, \emph{Delete}) to reconcile digital state with physical actuation (\Cref{fig:crutd}). Unlike conventional database CRUD, CRUTD primitives are specified with explicit pre-conditions and post-conditions, and (when coupled to actuation) are validated through device feedback and optional sensing to reduce state drift.

\begin{itemize}
    \item \textbf{Create:} Introduces a new entity into the system (e.g., registering an incoming reagent lot or issuing a new sample identifier). Logically, this adds a node into the resource tree. If the entity is not yet physically placed or localized, its pose is initialized as \texttt{unknown} until confirmed by actuation or sensing.
    \item \textbf{Update:} Modifies attributes of an existing entity without changing ownership/containment (e.g., updating a reactor setpoint, recording a measured pH value, marking an occupancy flag). Updates may be purely digital (annotation) or coupled to device execution (setpoint changes), depending on the attribute and policy.
    \item \textbf{Read:} Acquires state non-destructively (e.g., querying telemetry, validating vial presence via vision). Read operations do not modify logical containment or ownership and can serve as guards and validation steps in higher-level plans.
    \item \textbf{Transfer:} A first-class operation that changes containment/ownership by moving a resource from a source parent to a destination parent (e.g., moving liquid from \textit{SourceBottle} to \textit{DestVial}, or moving a vial between stations). Transfer is executed as an atomic transaction: the logical update is committed only after (i) actuation confirmation from the relevant devices and (ii) post-condition validation (e.g., sensor/vision confirmation where available).
    \item \textbf{Delete:} Removes an entity from the active schedulable set (e.g., disposing waste, archiving a sample). In practice, UniLabOS transfers the entity into a designated \textit{Trash}/\textit{Archive} subtree and marks it non-schedulable, preserving provenance while preventing further allocation.
\end{itemize}

\noindent\textbf{Transactional logic.}
For operations coupled to physical actuation (notably \emph{Transfer}), UniLabOS follows a two-phase reconcile pattern: (i) \emph{plan/validate} against the physical graph, resource constraints, and locks; (ii) \emph{execute/confirm} via device feedback and optional sensing. The host then commits the corresponding \texttt{ResourceDict} state update, or rolls back and emits a structured error event when post-conditions are not met. This yields a consistent sequence of CRUTD events that induces an execution provenance graph which can be queried and replayed.

\subsection{Dynamic driver registration via AST inspection}
\label{subsec:method-ast}

To support rapid hardware iteration, UniLabOS implements hot-swappable drivers with pre-instantiation validation based on static source inspection. Rather than relying on dynamic importing alone, UniLabOS parses driver source code using Python's AST prior to instantiation. The kernel extracts class metadata (method signatures, type hints, and docstrings) and checks conformance to expected A/R/A\&R conventions (e.g., \texttt{get\_}/\texttt{set\_} primitives and parameter schemas). Upon successful validation, the driver is registered without restarting the host process, and corresponding resource objects and capabilities are injected into the resource tree, reducing integration downtime and avoiding partial registrations.

\subsection{Semantic mapping to ROS~2 primitives}
\label{subsec:method-ros2-mapping}

UniLabOS maps driver methods to ROS~2 communication primitives according to operational semantics, enabling interoperability with robotics tooling and a shared message bus:

\begin{itemize}
    \item \textbf{State streaming (Topics):} High-frequency state publication (e.g., \texttt{get\_pose}, \texttt{get\_sensor\_data}) is mapped to ROS~2 topics. UniLabOS instantiates publishers that broadcast at a configured rate (typically 10--100\,Hz) to support monitoring and visualization without polling overhead.
    \item \textbf{Long-running execution (Actions):} Extended operations with progress and cancellability (e.g., \texttt{heat\_to}, \texttt{move\_to\_well}) are mapped to ROS~2 actions, providing goal--feedback--result semantics and cancellation.
    \item \textbf{Synchronous configuration (Services):} Request--response interactions that require explicit acknowledgment (e.g., configuration changes, admission checks, internal material validations) are mapped to ROS~2 services to standardize error reporting.
\end{itemize}

Upon registration, UniLabOS spawns the corresponding ROS nodes and binds them into a shared DDS domain, allowing chemical instruments and mobile robots to communicate through a uniform transport and introspection interface. Furthermore, to enable self-organizing enrollment, we implement a discovery and handshake protocol built on DDS discovery and ROS~2 services, where devices broadcast beacons and negotiate capabilities with the host before being admitted to the active resource tree.

\section{Acknowledgements}

We thank Tao Wang, Tengfei Yuan, Hua Han, and Cong Liu for developing the graphical user interface of UniLabOS. We are also grateful for insightful discussions with Wenjing Hong and Yikai Fang from IKKEM (Tan Kah Kee Innovation Laboratory, Xiamen University), Jian Jiang from ICCAS (Institute of Chemistry, Chinese Academy of Sciences), Rong Zhu from the College of Chemistry and Molecular Engineering, Peking University, and Chen Zhu from the Eastern Institute of Technology, Ningbo.

\bibliography{iclr2026_conference}
\bibliographystyle{iclr2026_conference}

\end{document}

%% file: math_commands.tex
\usepackage{amsmath,amsfonts,bm}

\def\eqref#1{equation~\ref{#1}}

\def\1{\bm{1}}

\DeclareMathAlphabet{\mathsfit}{\encodingdefault}{\sfdefault}{m}{sl}
\SetMathAlphabet{\mathsfit}{bold}{\encodingdefault}{\sfdefault}{bx}{n}

